\documentstyle[twocolumn,aps]{revtex}

\def\A{{\cal A}}\def\B{{\cal B}}
\def\H{{\cal H}}\def\R{{\cal R}}
\def\idty{{\rm 1\mkern -5.4mu I}}\def\tr#1{{\rm tr}\left(#1\right)}

\begin{document}

\title{EPR states for von Neumann algebras}
\author{R.~F. Werner }
\address{
Institut f\"ur Mathematische Physik, TU Braunschweig,
Mendelssohnstr.3, 38106 Braunschweig, Germany. \\ Electronic Mail:
r.werner@tu-bs.de}
\date{Oct. 16, 1999}

\maketitle

\begin{abstract}
In a recent paper [quant-ph/9910066],  Arens and Varadarajan gave
a characterization of what they call EPR-states on a bipartite
composite quantum system. By definition, such states imply perfect
correlation between suitable pairs of observables in the two
subsystems, and the task is to determine all such correlated pairs
for a given state. In this note the argument is shortened and
simplified, and  at the same time extended to observables in
general von Neumann algebras, which naturally arise in quantum
field theory.
\end{abstract}

\section{Introduction}

A key ingredient in the argument of the famous paper of Einstein
Podolsky and Rosen was the idea that in suitable states with
perfect correlations an ``element of reality'' of a subsystem
could be determined by measuring on a distant system, hence
without any perturbation. States with such perfect correlations
are nowadays used in many ways in Quantum Information Theory, and
even in practice. It was therefore interesting to see a paper
\cite{AV} in today's posting on quant-ph in which a mathematical
characterization of all such cases of perfect correlation was
undertaken. The present note arose from reading this paper, and
trying to find the key points in the rather cumbersome arguments.
Since this resulted in a much shorter argument applying to a wider
context, I compiled these notes for the benefit of other readers of the
archive.

\section{EPR states and Doubles}

We will look at the general situation of a quantum system, in
which two subsystems are singled out, whose observables are given
by two commuting von Neumann algebras $\A$ and $\B$, respectively.
That is, $\A$ is an algebra of bounded operators acting on a
Hilbert space $\H$, which is closed under limits in the weak
operator topology and the *-operation; the same holds for $\B$,
and any $A\in\A$ and $B\in\B$ commute. The special case considered
in \cite{AV} was the most familiar case, namely of a tensor
product Hilbert space $\H=\H_1\otimes\H_2,$ with $\A$ and $\B$ the
algebras of observables $A\otimes\idty$ and $\idty\otimes B$,
respectively. While this covers most situations considered in
quantum mechanics, and especially in quantum information theory
(see, however, \cite{Clif}),  this wider framework is needed in
quantum field theory and statistical mechanics of systems with
infinitely many degrees of freedom.

The key feature of the situation is that every observable
$A\in\A$ can be measured jointly with every $B\in\B$. Now in
\cite{AV} we find the following concept: a density operator $\rho$
on $\H$ is said to be an {\it EPR-state} for an observable
$A=A^*\in\A$, if there is an observable $A'\in\B$ such that the
joint distribution of $A$ and $A'$ with respect to the state is
concentrated on the diagonal\footnote{Actually, \cite{AV} consider
only vector states, and only require the existence of a
$\widetilde A'$ and a Borel function $g$ such that
$A'=g(\widetilde A')$ satisfies the above condition. But since we
may then just replace $A'$ by $g(\widetilde A')$, this only fakes
a gain in generality}. In other words, $A\in\A$ and $A'\in\B$ are
equal with probability one with respect to $\rho$, or,
\begin{equation}\label{double}
  \tr{\rho (A-A')^2}=0.
\end{equation}
We will call $A'$ the {\it double} of $A$ in $\B$, and denote by
$D(\A,\B,\rho)$ the subspace of elements $A\in\A$ for which a
double exists. This is the object determined in \cite{AV} in a
special case.

Now condition~(\ref{double}) can be written as $\tr{X^*X}=0$ with
$X=\sqrt\rho(A-A')$, hence implies $X=0$, or
\begin{equation}\label{rhoDa}
   \rho(A-A')=(A-A')\rho=0.
\end{equation}
Obviously, this equation makes sense also for non-hermitian
$A,A'$, so we use it to extend the definition of doubles and of
$D(\A,\B,\rho)$ to this case as well. Note that for vector states
$\rho=|\psi\rangle\langle\psi|$ this reduces to the two equations
$A\psi=A'\psi$ and $A^*\psi=A'^*\psi$.

If $A_1,A_2\in D(\A,\B,\rho)$, we have
$A_1A_2\rho=A_1A_2'\rho=A_2'A_1\rho=A_2'A_1'\rho$,  and similarly
on the other side, so $A_2'A_1'$ is a double of $A_1A_2$. This
makes $D(\A,\B,\rho)$ an algebra. Since we can choose the double
$A'$ to have the same norm as $A$ (truncate by a spectral
projection, if necessary. This won't make a difference on the
support of $\rho$) a simple compactness argument for weak limits
shows that $D(\A,\B,\rho)$ is also weakly closed, so it is a von
Neumann algebra.

To further identify this algebra note that, for $A\in D(\A,\B,\rho)$
and any
$A_1\in\A$, $\tr{\rho AA_1}=\tr{\rho A'A_1}=\tr{\rho
A_1A'}=\tr{A'\rho A_1}=\tr{A\rho A_1}=\tr{\rho A_1A}$. That is to
say $D(\A,\B,\rho)$ is contained in the {\it centralizer} of
$\rho$ in $\A$, which we will denote by $C_\rho(\A)$. Note that
the centralizer does not depend on the entire density operator
$\rho$, but only on the linear functional it induces on $\A$. So
in the special case when $\A$ is isomorphic to the bounded
operators on a Hilbert space $\H_A$, we can express this restriction by
a
density operator\footnote{In \cite{AV} this density operator is
written as $\rho_A=L_\psi^*L_\psi$, where $\psi\in\H_A\otimes\H_B$
is the vector determining $\rho$, and $L_\psi:\H_A\to\H_B$ is the
conjugate linear Hilbert-Schmidt operator they could have defined
in a basis free way through the formula
$\langle\psi,\chi_A\otimes\chi_B\rangle=\langle
L_\psi(\chi_A),\chi_B\rangle$ and an invocation of Riesz's
Theorem. } $\rho_A$ on $\H_A$. The centralizer in this case is simply
the
set of operators commuting with $\rho_A$.

In the trivial case considered in \cite{AV} it is easy to see
that, conversely, any element of the centralizer indeed has a
double. In the more general situation that is not true, but there
is one standard situation in which it is. Moreover, the general
case can be understood completely in terms of the standard case.
In this standard case $\rho$ is a vector state, given by a vector
$\psi$, which is cyclic and separating for $\A$, i.e., $\A\psi$ is
dense in $\H$, and $A\psi=0$ for $A\in\A$ implies $A=0$. In this
situation the modular theory of Tomita and Takesaki applies, and
we get the following Theorem:

\noindent{\bf Theorem}{\it Let $\A$ be a von Neumann algebra with
cyclic and separating vector $\psi$, and set
$\rho=|\psi\rangle\langle\psi|$. Then
\[D(\A,\A',\rho)=C_\rho(\A).\]
Moreover, the double $A'\in\A'$ of
any $A\in C_\rho(\A)$ is unique.}

The following proof is sketchy, because it fails to explain
modular theory, which is, however, well documented and accessible
(e.g., \cite{BraRo}). The basic object of that theory is the
unbounded conjugate linear operator $S$ defined by
$SA\psi=A^*\psi$. Its polar decomposition $S=J\Delta^{1/2}$ yields
an antiunitary involution $J$ such that $J\A J=\A'$. Then $A\in\A$
belongs to the centralizer iff $\Delta$ commutes with $A$ in the
sense that $\Delta^{it} A\Delta^{-it}=A$, which also implies
$\Delta A\psi=A\psi$ and $\Delta A^*\psi=\psi$. We claim that in
that case $A'=JA^*J\in\A'$ is a double of $A$ in $\A'$: we have
$A'\psi=JA^*J\psi=JA^*\psi=JSA\psi=\Delta A\psi=A\psi$. For the
uniqueness of the double we only need that $\psi$ is cyclic, which
is equivalent to $\psi$ being separating for $\A'$. Then any two
doubles $A'$ and $\widetilde A'$, which have to satisfy
$A'\psi=A\psi=\widetilde A'\psi$ must be equal. This concludes the
proof.

As a corollary we can compute the algebra: $D(\A,\B,\rho)$ for
$\B\subset\A'$. Since a double in $\B$ is also a double in $\A'$,
it is the subalgebra of $C_\rho(\A)$ for which the doubles $JA^*J$
lie in $\B$. That is,
\begin{equation}\label{DAB}
   D(\A,\B,\rho)=C_\rho(\A)\cap J\B J.
\end{equation}

To reduce the general case to the case with cyclic and separating
vector for $\A$, one first enlarges the Hilbert space by a
suitable tensor factor, so that  $\rho$ extends to a pure state
$|\psi\rangle\langle\psi|$ on the enlarged space. Denote by $\R$
and $\R'$ the closed subspaces generated by $\A\psi$ and
$\A'\psi$, respectively. Then, for $A\in\A$, we have
$A\R\subset\R$, and if $A$ has a double in $\A'$, we get
$AB'\psi=B'A\psi=B'A'\psi\in\R'$, which implies that
$A\R'\subset\R'$. The same arguments apply to the equation
$A^*\psi=A'^*\psi$, so we find that both $A$ and its double $A'$
have to commute with both the projection $R\in\A'$ onto $\R'$ and
the projection $R'\in\A$ onto $\R'$.

Hence any $A\in D(\A,\B,\rho)$ can be split in $\A$ into
$A=(\idty-R')A(\idty-R')+R'AR'$, where the first summand has zero
as its double, and only the second summand is of interest in this
problem. Similarly, any putative double can be split into an
irrelevant part $(\idty-R)A'(\idty-R)$, which only creates
non-uniqueness, and an essential part $RA'R$. Hence we may
restrict consideration to the subspace $\R\cap\R'$ on which $\psi$
is indeed cyclic and separating.

\section{Concluding Remarks}

Finally, a comment seems in order about the relevance of the
generalization of the concept of EPR-states to the general von
Neumann algebraic setting. First of all, in quantum field theory,
type I algebras (in von Neumann's classification; i.e., those
considered in \cite{AV}) never appear as the observable algebras
of local regions, but interesting insights can be gained from
studying and EPR-phenomena where spacelike separated are localized
close to each other (see \cite{SW} and references therein).

Secondly, there is a conclusion in \cite{AV}, which may seem
striking at first glance, namely that an observable which
possesses a double necessarily has discrete spectrum. In view of
the present note this becomes immediately clear: it is an artefact
of the type~I situation, where all centralizers are sums of finite
dimensional matrix algebras. As soon as one drops this constraint,
the conclusion disappears: a prototype is the trace on a type
II$_1$ factor, where the centralizer is the whole algebra, and
many observables with continuous spectrum exist. In fact, such an
algebra, which arises as the tensor product of infinitely many
qubit pairs with maximal violations of Bell's inequality, plays a
canonical role in the study  of extremely strong violations of
Bell's inequalities in \cite{SW}.


\begin{references}

\bibitem{AV} R. Arens and V.S. Varadarajan, {\it On the concept of
EPR states and their structure}, quant-ph/9910066.

\bibitem{Clif} R. Clifton, H. Halvorson, and A. Kent:
{\it Non-local Correlations are Generic in Infinite-Dimensional
Bipartite Systems}, quant-ph/9909016

\bibitem{BraRo} O. Bratteli and D.W. Robinson,
{\it Operator algebras and quantum statistical mechanics}, Vol.1,
Springer, 1979/1987

\bibitem{SW} S.J. Summers and R.F. Werner,
``On Bell's inequality and algebraic invariants'',
     {\it Lett.Math.Phys.}{\bf 33}(1995) 321--334

\end{references}
\end{document}